\documentclass{article}
\usepackage{spconf,amsmath,epsfig,amssymb}

\newcommand{\alphab}{{\mbox{\boldmath $\alpha$}}}

\newcommand{\sbb}{{\bf s}}
\newcommand{\halphab}{{\mbox{\boldmath $\hat \alpha$}}}
\newcommand{\htheta}{\hat \theta}

\newcommand{\Sbb}{{\bf s^{\cal{(F)}}}}

\newcommand{\Wb}{{\bf W}}
\newcommand{\Mb}{{\bf M}}

\newcommand{\phib}{{\mbox{\boldmath $\varphi$}}}
\newcommand{\PHIb}{{\mbox{\boldmath {$\Phi$ }}}}

\newcommand{\bPHIb}{{\mbox{\boldmath  $\bar \Phi$}}}
\newcommand{\argmin}{\mathop{\textrm{argmin}}}

\title{A first step to Convolutive Sparse Representation}
\twoauthors
   {
    H. Firouzi,
    M. Babaie-Zadeh\sthanks{This work has been partially funded
    by Sharif University of Technology, ISMO and Iran NSF (INSF).},
    A. Ghasemian Sahebi}
   {Sharif University of Technology\\
   Department Of Electrical Engineering\\
   Tehran, Iran}
   {C. Jutten}
   {Institut National Polytechnique\\ de Grenoble (INPG),
   Labratoire \\des Images et de Signaux (LIS)\\
   Grenoble, France}

\begin{document}
%
\maketitle
\begin{abstract}

In this paper an extension of the sparse decomposition problem is
considered and an algorithm for solving it is presented. In this
extension, it is known that one of the shifted versions of a
signal $\sbb$ (not necessarily the original signal itself) has a
sparse representation on an overcomplete dictionary, and we are
looking for the sparsest representation among the representations
of all the shifted versions of $\sbb$. Then, the proposed
algorithm finds simultaneously the amount of the required shift,
and the sparse representation. Experimental results emphasize on
the performance of our algorithm.
\end{abstract}

\section{Introduction}
\label{sec:intro}

In the classical atomic decomposition problem~\cite{ChenDS99}, we
have a signal $s(t)$ whose samples are collected in the $n \times
1$ signal vector $\sbb=[s(1),\dots,s(n)]^T$ and we would like to
represent it as a linear combination of $m$, $n \times 1$ signal
vectors $\{\phib_{i}\}_{i=1}^{m}$. After~\cite{MallZ93}, the
vectors $\phib_{i}$, $1 \leq i \leq m$ are called \emph{atoms}
and they collectively form a \emph{dictionary} over which the
signal is to be decomposed. We may write
\begin{equation}\label{eq:rep}
\sbb =\sum_{i=1}^{m} \alpha_{i} \phib_{i} = \PHIb \alphab,
\end{equation}
where $\PHIb \triangleq [\phib_{1},\dots,\phib_{m}]$ is the $n
\times m$ dictionary (matrix) and $\alphab \triangleq
(\alpha_1,\dots,\alpha_m)^T$ is the $m \times 1$ vector of
coefficients. A dictionary with $m > n$ is called
\emph{overcomplete}. Although, $m=n$ is sufficient to obtain such
a decomposition (like what is done in Discrete Fourier Transform),
using overcomplete dictionaries has a lot of advantages in many
diverse applications (refer for example to~\cite{DonoET06} and the
references in it). Note that for the overcomplete case, the
representation is not unique, but all these applications need a
sparse representation, that is, the signal $\sbb$ should be
represented as a linear combination of as small as possible
number of atoms of the dictionary. It has been
shown~\cite{Dono04,GoroR97} that with some mild conditions on the
dictionary matrix, if there is a sparse representation with at
most $n/2$ non-zero coefficients, then this representation is
unique. The main approaches for finding this sparse solution
include Matching Pursuit (MP)~\cite{MallZ93,KrstG06},
FOCUSS~\cite{GoroR97}, Basis Pursuit (BP)~\cite{ChenDS99}, and
Smoothed $\ell^0$ (SL0)~\cite{MohiBJ07}.

In this paper, we introduce a generalization of this classical
problem to the case that we call `convolutive sparse representation'. In this
case, it is known that the signal $\sbb$ has a sparse
representation not over the dictionary itself, but over some
(unknown) shifted versions of the atoms. To state the problem
more clearly, consider a representation of the form:
\begin{equation}\label{eq:defeq}
\sbb=\sum_{i=1}^{m}\alpha_i{{\phib}_i}^{(k_i)},
\end{equation}
where ${{\phib}_i}^{(k_i)}$ stands for the $k_i$-sample
(circularly) shifted version of $\phib_i$. Then, our problem is to
find the sparsest representation in the form of (\ref{eq:defeq})
among all the possible values for $k_1, \dots, k_m$.

Note also that the Fourier transform does not convert this
problem to the classical sparse representation (\ref{eq:rep}) in
the frequency domain: The problem in the transformed domain will
be similar to (\ref{eq:rep}), but with {\em time varying
$\alpha_i$'s}.

In this paper, we address only a special case of the general
problem (\ref{eq:defeq}), that is, where all the shifts $k_i$ are
equal. This is equivalent to this simplified problem: an unknown
shifted version of $\sbb$ has a sparse representation over the
dictionary, and we would like to find this representation.

One of the trivial applications of the general problem is to
reduce the size of the dictionary in atomic decomposition
applications. An example for the applications of the above
simplified problem is where our recorded signal, which has to be
decomposed as a combination of a small number of atoms of the
dictionary, is shifted relative to its underlying atoms that
already exist in the dictionary.

The paper organized as follows. In Section~\ref{sec:Main} the
main idea of the algorithm is introduced. The resulting algorithm
is then stated in Section~\ref{sec:FinalAlgorithm}. Finally,
simulation results of the algorithm are presented in
Section~\ref{sec:simulations}.

\section{Main Idea}
\label{sec:Main} Consider a dictionary with atoms
${\phib}_1,{\phib}_2,\cdots,{\phib}_m$. The problem is then to
sparsely decompose an $n\times 1$ vector\, $\sbb$ as a linear
combination of shifted atoms of the dictionary (in this paper, the
shifts are assumed to be circular). One trivial solution to the
problem is to insert all shifted atoms in the dictionary and then
find the sparsest representation of the vector $\sbb$ for that
dictionary using the conventional atomic decomposition methods.
However, this direct solution demands a high computational and
storage load.

Let also that $k_i$ be a continuous variable (a non-integer shift
$k_i$ can be imagined as shifting the hull of signal and then
re-sampling it). For handling circular shifts more easily, we
take the Discrete Fourier Transform (DFT) of both sides of
(\ref{eq:defeq}) to obtain:
\begin{equation}
\label{eq:XW} \Sbb=\sum_{i=1}^{m}\alpha_i
\Wb_i{\phib}_{i}^{\cal{(F)}}
\end{equation}
in which $\Sbb$ and $\phib_{i}^{\cal{(F)}}$ are the DFTs of the
signals $\sbb$ and $\phib_i$ repectively, and
\begin{displaymath}
\Wb_i \triangleq \left(\begin{array}{c c c c c}
{w_i}^0 & 0 & 0 & 0 & 0\\
0 & {w_i}^1 & 0 & 0 & 0\\
0 & 0 & {w_i}^2 & 0 & 0\\
0 & 0 & 0 & \ddots  & 0\\
0 & 0 & 0 & 0 & {w_i}^{n-1}
\end{array}\right), w_i \triangleq e^{-j\frac{2\pi}{n}k_i}
\end{displaymath}

As stated in the introduction, in this paper we consider only the
case in which $w_1=w_2=\cdots =w_m$, and we present an iterative
algorithm to solve the problem in this case. This case is
equivalent to assuming that the atoms of the dictionary are fixed
and the signal $\sbb$ is shifted in opposite direction. In this
case:
\begin{displaymath}
\Wb_1=\Wb_2=\cdots=\Wb_m=\Wb,
\end{displaymath}
and hence from \eqref{eq:XW} we have:
\begin{equation}
\label{eq:eqshift} \Sbb=\Wb \sum_{i=1}^{m}\alpha_i
\phib_{i}^{\cal{(F)}}
\end{equation}
or:
\begin{equation}
\label{eq:WpX} \Wb^\prime \Sbb=\sum_{i=1}^{m}\alpha_i
\phib_{i}^{\cal{(F)}}=\PHIb^{\cal{(F)}}\alphab
\end{equation}
where:
\begin{displaymath}
\Wb^\prime=\Wb^{-1}=\left(\begin{array}{c c c c c}
{{w^\prime}}^0 & 0 & 0 & 0 & 0 \\
0 & {{w^\prime}}^1 & 0 & 0 & 0\\
0 & 0 & {{w^\prime}}^2 & 0 & 0\\
0 & 0 & 0 & \ddots  & 0\\
0 & 0 & 0 & 0 & {{w^\prime}}^{n-1}
\end{array}\right)\\
\end{displaymath}
in which ${w^\prime}=e^{j\frac{2\pi}{n}k}$. Now the problem is to
find the sparsest solution of \eqref{eq:WpX}. To do so, we should
have some criterion $F(\alphab)$ for sparseness of the solution
vector $\alphab$ and optimize that criterion subject to the
constraint \eqref{eq:WpX} using optimization methods. Note also
that one of the unknows, $k$ does not exist in the objective
function $F(\alphab)$, and appears only in the constraint
\eqref{eq:WpX}. As their objective functions, two classical
sparse decomposition approaches use
$\ell^1$-norm~\cite{ChenDS99}, and smoothed $\ell^0$ (SL0)
norm~\cite{MohiBJ07}. Here we use the second one, because it
results in a very fast and accurate algorithm in classical atomic
decomposition~\cite{MohiBJ07}, and also because it is a
differentiable measure of the sparsity of $\alphab$. Smoothed
$\ell^0$-norm of a vector is an approximation to its $\ell^0$-norm
(number of its non-zero element), and is defined as:
\begin{equation}
\label{eq:sparsity}
F(\alphab)=m-\sum_{i=1}^m{e^{-{\alpha_i}^2/{2\sigma^2}}}
\end{equation}
where $\sigma$ is a parameter which specifies a tradeoff between
smoothness and the accuracy of approximation: the smaller
$\sigma$, the better approximation of the $\ell^0$ norm; the
larger $\sigma$, the smoother objective function.

On the other hand, \eqref{eq:WpX} can be written as:
\begin{equation}
\label{eq:Subject1} G(\alphab,w')=\|\Wb^\prime
\Sbb-\PHIb^{\cal{(F)}} \alphab\|^2=0
\end{equation}
Now we should minimize \eqref{eq:sparsity} subject to the
\eqref{eq:Subject1}, for a small value of $\sigma$. Note that one
of the optimization variables ($w'$), is not present in
$F(\alphab)$, and appears only in \eqref{eq:Subject1}.

Note that for small values of $\sigma$, $F$ contains a lot of
local minima. Consequently, it is very difficult to directly
minimize this function for very small values of $\sigma$. The idea
of~\cite{MohiBJ07} for escaping from local minima is then to
decrease the value of $\sigma$ gradually: for each value of
$\sigma$ the minimization algorithm is initiated with the
minimizer of the $F$ for the previous (larger) value of
$\sigma$.  This idea of minimizing a non-convex function is
called Graduated Non-Convexity~\cite{BlakZ87}, and is also used
in simulated annealing methods.

To start the minimization, we should find a proper initial guess
for the solution $\alphab_0$, that is, the initial estimation of
the sparsest solution of $\PHIb\alphab=\sbb$. It has been shown
that for the case of the simple sparse decomposition, the best
initial value for $\alphab$ is the minimum $\ell^2$-norm solution
of $\PHIb\alphab=\sbb$, that is, $\alphab_0=\PHIb^T (\PHIb
\PHIb^T)^{-1}\sbb$~\cite{MohiBJ07}. The reason is that this
solution minimizes the function $F(\alphab)$ subject to
$\PHIb\alphab=\sbb$ where $\sigma$ goes to infinity. Despite the
fact that our method is somehow different with the method
presented in~\cite{MohiBJ07}, we use the same initialization for
our algorithm. Since we also have the variable $w'$, we should
start from the sparsest $\alphab(k)=\PHIb^T
(\PHIb\PHIb^T)^{-1}\sbb^{(k)}$ vector. Let $k_0 = \argmin_k
F(\alphab(k))$, for $k=1,2,\dots,n$. Then we choose
$\alphab(k_0)$ to be the starting point of our algorithm.

Because of noise, if our algorithm tries to satisfy
\eqref{eq:Subject1} exactly, it would be very sensitive to noise.
Consequently, we try to satisfy this equation approximately. We
realize this idea by minimizing the function $H$ defined below
with respect to $\alphab$ and ${w^\prime}$:
\begin{equation}\label{eq:H}
H(\alphab,w')=\lambda G(\alphab,w')+(1-\lambda)F(\alphab)
\end{equation}
where $0<\lambda<1$ is a constant that specifies the weight which
is given to satisfying \eqref{eq:Subject1}. This equation can be
interpreted as a trade-off between the accuracy of the
decomposition and maximizing the sparsity.

By letting $w'=e^{j\theta}$, the final objective function
$H(\alphab,\theta)$ will be a real-valued function of real-valued
variables $\alphab$ and $\theta$. For each $\sigma$, this
function may be minimized by gradient based algorithms
(specifically steepest descent). Direct calculations show:
\begin{gather}
\label{eq:Galpha}
\begin{split}
\frac{\partial H}{\partial \alphab} &= 2 \lambda
\Re\{\PHIb^{\cal{(F)}\mathrm{T}}(\PHIb^{\cal{(F)}}\alphab-\Wb'\Sbb)\} + \\
& (1-\lambda)(1/\sigma^2)[\alpha_{1}e^{(-\alpha_{1}^{2}/2\sigma^{2})},
\dots,\alpha_{m}e^{(-\alpha_{m}^{2}/2\sigma^{2})}]^{T}
\end{split} \\
\label{eq:Gtheta}
\frac{\partial H}{\partial \theta}=-2\Re\{\Sbb\Mb\Wb'\bPHIb\}
\end{gather}
where $\Mb \triangleq \mathrm{diag} (0,1j,\dots,(n-1)j)$.


\section{The Final Algorithm}
\label{sec:FinalAlgorithm} The final algorithm of the proposed
method is given in Fig.~\ref{fig:algorithm}. As seen in the
algorithm, the final values of the previous estimation are used
for initialization of the next steepest descent. As explained in
the previous section, the decreasing sequence of $\sigma$ is used
to escape from getting trapped into local minima.

In the minimization part, the steepest descent with variable
step-size ($\mu$) has been used: If $\mu$ is such that
$H(\alphab-\mu \frac{\partial H}{\partial
\alphab},\theta-\mu\frac{\partial H}{\partial
\theta})<H(\alphab,\theta)$ we multiply the value of $\mu$ by 1.2
for the next iteration. Otherwise if $\mu$ is such that
$H(\alphab-\mu\frac{\partial H}{\partial
\alphab},\theta-\mu\frac{\partial H}{\partial \theta})\ge
H(\alphab,\theta)$ we multiply the value of $\mu$ by 0.5 for the
next iteration.

\begin{figure}[!h]
\small

\vrule
\begin{minipage}{8.5cm} 
\hrule \vspace{0.5em} 
\begin{minipage}{7.5cm} 
        \begin{itemize}
        \item Initialization:

           \begin{enumerate}

             \item{Let: $F(\alphab)=m-\sum_{i=1}^m{e^{-{\alpha_i}^2/{2\sigma^2}}}$ and
             $\alphab(k)=\PHIb^T (\PHIb\PHIb^T)^{-1}\sbb^{(k)}$}.

             \item Find the minimum of $F(\alphab(k))$ for $k=1, 2,\cdots,n$. Assuming
             this minimum occurs for $k=k_0$, let $\halphab_0=\alphab(k_0)$ and
             $\htheta_0=\frac{2\pi}{n}k_0$.

             \item Choose a suitable decreasing sequence for
             $\sigma=[\sigma_{1}\ldots\sigma_{R}]$. Choose a
             small value for $\mu$.

             \item Let $\Mb= \mathrm{diag} (0,1j,\dots,(n-1)j)$.

           \end{enumerate}

        \item For $r=1,\dots,R$:
          \begin{enumerate}
             \item Let $\sigma=\sigma_r$.
             \item Minimize (approximately) the function
             $H(\alphab,\theta)$
             using $L$ iterations of the steepest descent
             algorithm:
             \begin{itemize}
               \item Initialization: $\alphab=\halphab_{r-1}$ and $\theta=\hat \theta_{r-1}$.
               \item for $l=1\dots L$ (loop $L$ times):
                    \begin{enumerate}
                       \item Calculate $\frac{\partial H}{\partial \alphab}$ and $\frac{\partial H}{\partial \theta}$
                       from (\ref{eq:Galpha}) and (\ref{eq:Gtheta}), respectively.

                       \item If
                       $H(\alphab-\mu\frac{\partial H}{\partial \alphab},\theta-\mu\frac{\partial H}{\partial \alphab})<H(\alphab,\theta)$
                       let $\rho=1.2$ else $\rho=0.5$.

                       \item Let $\alphab\leftarrow \alphab-\mu {\partial H}/{\partial \alphab}$\\
                       and $\theta\leftarrow \theta-\mu {\partial H}/{\partial \theta}$.

                       \item Let $\mu\leftarrow \mu\times\rho$.

                    \end{enumerate}
             \end{itemize}

             \item Set $\halphab_r=\alphab$ and $\htheta_r=\theta$.
          \end{enumerate}
        \item Let $\halphab=\halphab_R$ and $\htheta=\htheta_R$. The final
        coefficient vector is $\halphab$ and the final shift value is
        $n\times\frac{\htheta}{2\pi}$.
        \end{itemize}
\end{minipage}
\vspace{0.5em} \hrule
\end{minipage}\vrule \\
\\
\caption{The final algorithm} \label{fig:algorithm}
\end{figure}

\section{EXPERIMENTAL RESULTS}
\label{sec:simulations} In order to experimetally evaluate our method, we
generated a random dictionary $\PHIb$ which had $80$ atoms and
each atom was a signal of length $40$ (thus we assumed $m=80$ and
$n=40$ in our simulations). Then we created a synthetic vector
$\sbb$ by generating a sparse coefficient vector $\alphab$ at
random, using a Bernoulli-Gaussian model: each coefficient is
`active' with probability $p$, and is `inactive' with probability
$1-p$. If it is active, its value is modeled by a zero-mean
Gaussian random variable with variance $\sigma^2_\mathrm{on}$; if
it is not active, its value is modeled by a zero-mean Gaussian
random variable with variance $\sigma^2_\mathrm{off}$, where
$\sigma^2_\mathrm{off} \ll \sigma^2_\mathrm{on}$. Consequently,
each $\alpha_i$ is distributed as:
\begin{equation}
\label{eq: the sources model} \alpha_{i}\sim p \cdot
\mathcal{N}(0,\sigma_{\mathrm{on}})+(1-p) \cdot
\mathcal{N}(0,\sigma_{\mathrm{off}}),
\end{equation}
where $p$ denotes the probability of activity of the coefficient,
and sparsity implies that $p \ll 1$. 
In our simulations we have fixed $\sigma_{\mathrm{on}}=1,
\sigma_{\mathrm{off}}=0.01$, $p=0.1$, and $\lambda=0.75$.

Then we created the signal $\sbb$ by $\sbb=\PHIb \alphab+{\bf n}$, where $\bf n$
is an additive white Gaussian noise with zero mean and standard deviation $\sigma_{\mathrm{n}}=0.01$.
Finally, we shifted the $\sbb$ vector circularly by $k$
samples where $k$ was a random number from $0$ to $39$.
We applied our algorithm to convolutively decompose this vector
$\sbb$ over the dictionary $\PHIb$.

\begin{figure}[htb]
  \centerline{\epsfig{figure=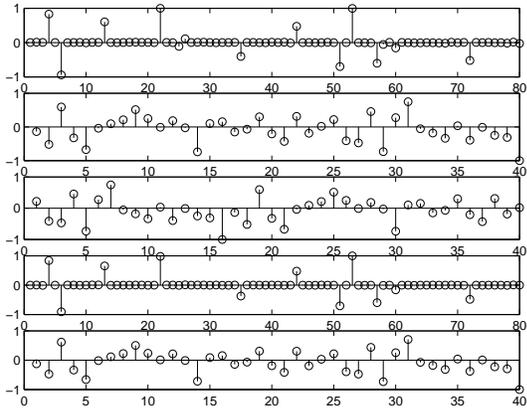,width=8.5cm}}
  \vspace{0.7cm}
  \caption{A sample of our experiments. From top to bottom,
  first plot represents a randomly generated coefficient vector $\alphab$,
  second plot is the synthetic vector $\sbb$ which has the coefficient
  vector $\alphab$ on the randomly generated dictionary $\PHIb$, third plot
  is a randomly shifted version of vector $\sbb$ which is the input of our algorithm,
  fourth plot is the estimated coefficient vector $\halphab$, and the last plot is the
  vector $\hat \sbb$ which has the coefficient vector $\halphab$.}
\label{fig:sample}
\end{figure}

The simulation was repeated 1000 times with randomly generated coefficients, dictionary and
the shift of the signal,
and it was seen that in 992 experiment the algorithm could sucessfully estimate
the shift value and the coefficient
vector $\alphab$. In average, the Signal to Noise
Ratio\footnote{Signal to Noise Ratio is defined as
$10\log_{10}\frac{\Vert \alphab \Vert^{2}}{\Vert \halphab-\alphab
\Vert^{2}} $ where $\halphab$ is the estimated coefficient
vector.} (SNR) was greater than 24dB. Figure \ref{fig:sample} shows one
of the runs of these experiments. In the other
8 experiments the algorithm felled into local minima, and could not correctly estimated $\alphab$
and $\theta$.

\begin{figure}[htb]
  \centerline{\epsfig{figure=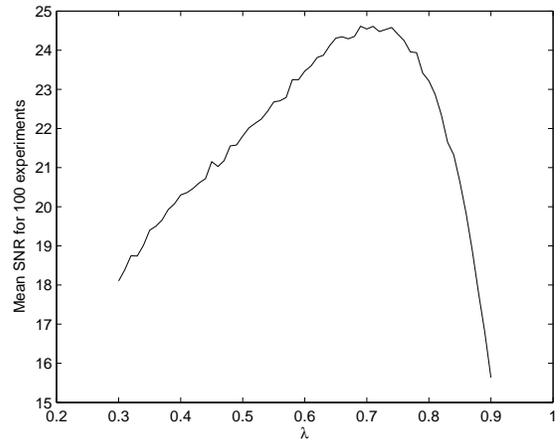,width=8.5cm}}
  \vspace{0.7cm}
  \caption{Output SNR versus $\lambda$.}
\label{fig:lambda}
\end{figure}

In order to see the effect of $\lambda$ on the estimation quality,
the algorithm was repeated for $\lambda$'s between 0.3 and 0.9
(outside this interval SNR decreases rapidly). For each value of
$\lambda$ we repeated the algorithm 100 times and the mean SNR for
each $\lambda$ is computed. The mean SNR is plotted versus
$\lambda$ in Fig.~\ref{fig:lambda}.

\section{CONCLUSION}

In this paper, a new method was proposed as the first step for
solving the convolutive sparse decomposition problem. The
proposed method can be used in the cases in which we know that
one of the shifted versions of a signal $\sbb$ has a sparse
representation on an overcomplete dictionary, and we are looking
for the sparsest representation among the representations of all
the shifted versions of $\sbb$. We used Discrete Fourier
Transform (DFT) to convert the problem to a continuous
optimization problem. The proposed method was fast because of
using the idea of smoothed $\ell^0$-norm~\cite{MohiBJ07}.
Experimental results emphasized on the performance of the
proposed algorithm.

It seems that the proposed algorithm can be generalized for
applying to the general convolutive sparse representation problem
(in which the shift values $k_i$ are not necessarily equal).
However, our simulations show that the main difficulty of such a generalization
is that the algorithm very oftenly traps into local minima. Such a generalization
is currently under study in our group.


\bibliography{SepSrc}

\begin{thebibliography}{1}

\bibitem{ChenDS99}
S.~S. Chen, D.~L. Donoho, and M.~A. Saunders,
\newblock ``Atomic decomposition by basis pursuit,''
\newblock {\em SIAM Journal on Scientific Computing}, vol. 20, no. 1, pp.
  33--61, 1999.

\bibitem{MallZ93}
S.~Mallat and Z.~Zhang,
\newblock ``Matching pursuits with time-frequency dictionaries,''
\newblock {\em IEEE Trans. on Signal Proc.}, vol. 41, no. 12, pp. 3397--3415,
  1993.

\bibitem{DonoET06}
D.~L. Donoho, M.~Elad, and V.~Temlyakov,
\newblock ``Stable recovery of sparse overcomplete representations in the
  presence of noise,''
\newblock {\em IEEE Trans. Info. Theory}, vol. 52, no. 1, pp. 6--18, Jan 2006.

\bibitem{Dono04}
D.~L. Donoho,
\newblock ``For most large underdetermined systems of linear equations the
  minimal $l^{1}$-norm solution is also the sparsest solution,''
\newblock Tech. {R}ep., 2004.

\bibitem{GoroR97}
I.~F. Gorodnitsky and B.~D. Rao,
\newblock ``Sparse signal reconstruction from limited data using {FOCUSS}, a
  re-weighted minimum norm algorithm,''
\newblock {\em IEEE Transactions on Signal Processing}, vol. 45, no. 3, pp.
  600--616, March 1997.

\bibitem{KrstG06}
S.~Krstulovic and R.~Gribonval,
\newblock ``\mbox{MPTK}: Matching pursuit made tractable,''
\newblock in {\em ICASSP'06}, 2006.

\bibitem{MohiBJ07}
G.H. Mohimani, M.~Babaie-Zadeh, and C.~Jutten,
\newblock ``Fast sparse representation based on smoothed $l^0$-norm,''
\newblock in {\em ICA2007}, London, September 2007.

\bibitem{BlakZ87}
A.~Blake and A.~Zisserman,
\newblock {\em Visual Reconstruction},
\newblock MIT Press, 1987.

\end{thebibliography}
\bibliographystyle{IEEEbib}
\end{document}